\documentclass[twocolumn]{aastex631}

\begin{document}

\shortauthors{Luhman et al.}

\shorttitle{JWST/NIRSpec Observations of TWA~27B}

\title{JWST/NIRSpec Observations of the Planetary Mass Companion 
TWA~27B\footnote{Based on observations made with the NASA/ESA/CSA 
James Webb Space Telescope, the Gaia mission, the Two Micron All Sky Survey, 
and the Wide-field Infrared Survey Explorer.}}

\author{K. L. Luhman}
\affiliation{Department of Astronomy and Astrophysics,
The Pennsylvania State University, University Park, PA 16802, USA;
kll207@psu.edu}
\affiliation{Center for Exoplanets and Habitable Worlds, The
Pennsylvania State University, University Park, PA 16802, USA}

\author{P. Tremblin}
\affiliation{Universit\'{e} Paris-Saclay, UVSQ, CNRS, CEA, 
Maison de la Simulation, 91191, Gif-sur-Yvette, France}

\author{S. M. Birkmann}
\affiliation{European Space Agency (ESA), ESA Office, Space Telescope
Science Institute, 3700 San Martin Drive, Baltimore, MD 21218, USA}

\author{E. Manjavacas}
\affiliation{AURA for the European Space Agency, Space Telescope Science
Institute, 3700 San Martin Drive, Baltimore, MD 21218, USA}
\affiliation{Department of Physics \& Astronomy, Johns Hopkins University, 
Baltimore, MD 21218, USA}

\author{J. Valenti}
\affiliation{Space Telescope Science Institute, 3700 San Martin Drive, 
Baltimore, MD 21218, USA}

\author{C. Alves de Oliveira}
\affiliation{European Space Agency, European Space Astronomy Centre,
Camino Bajo del Castillo s/n, 28692 Villanueva de la Ca\~{n}ada, Madrid, Spain}

\author{T. L. Beck}
\affiliation{Space Telescope Science Institute, 3700 San Martin Drive, 
Baltimore, MD 21218, USA}

\author{G. Giardino}
\affiliation{ATG Europe for the European Space Agency, ESTEC, Noordwijk, 
The Netherlands}

\author{N. L\"{u}tzgendorf}
\affiliation{European Space Agency (ESA), ESA Office, Space Telescope
Science Institute, 3700 San Martin Drive, Baltimore, MD 21218, USA}

\author{B. J. Rauscher}
\affiliation{NASA Goddard Space Flight Center, Observational Cosmology
Laboratory, Greenbelt, USA}

\author{M. Sirianni}
\affiliation{European Space Agency (ESA), ESA Office, Space Telescope
Science Institute, 3700 San Martin Drive, Baltimore, MD 21218, USA}

\begin{abstract}

We present 1--5~\micron\ spectroscopy of the young planetary mass companion 
TWA~27B (2M1207B) performed with NIRSpec on board the James Webb Space 
Telescope. In these data, the fundamental band of CH$_4$ is absent and 
the fundamental band of CO is weak. 
The nondetection of CH$_4$ reinforces a previously observed trend of weaker
CH$_4$ with younger ages among L dwarfs, which has been attributed to
enhanced non-equilibrium chemistry among young objects.  
The weakness of CO may reflect an additional atmospheric property that 
varies with age, such as the temperature gradient or cloud thickness.
We are able to reproduce the broad shape of the spectrum with an {\tt ATMO}
cloudless model that has $T_{\rm eff}=1300$~K, non-equilibrium chemistry, 
and a temperature gradient reduction caused by fingering convection. 
However, the fundamental bands of CH$_4$ and CO are somewhat stronger in
the model. In addition, the model temperature of 1300~K is higher than
expected from evolutionary models given the luminosity and age of TWA~27B
($T_{\rm eff}=1200$~K).
Previous models of young L-type objects suggest that the inclusion of clouds
could potentially resolve these issues; it remains to be seen whether cloudy
models can provide a good fit to the 1--5~\micron\ data from NIRSpec.
TWA~27B exhibits emission in Paschen transitions and the He~I triplet 
at 1.083~\micron, which are signatures of accretion that provide
the first evidence of a circumstellar disk.
We have used the NIRSpec data to estimate the bolometric luminosity of TWA~27B
(log~$L/L_\odot=-4.466\pm0.014$), which implies a mass of
5--6~$M_{\rm Jup}$ according to evolutionary models.

\end{abstract}

\section{Introduction}
\label{sec:intro}

With an age of $\sim$10~Myr \citep{bar06,bel15},
the TW Hya association (TWA) is the youngest known stellar population within
100~pc \citep{del89,gre92,kas97,web99,gag17,luh23twa}.
Because of its youth and proximity, \citet{giz02} performed a survey
for brown dwarfs in TWA using photometry from the Two Micron All Sky
Survey shortly after its completion \citep[2MASS,][]{skr06}.
Two substellar members of TWA were discovered through that work,
2MASSW J113951$-$3159211 and 2MASSW J1207334$-$393254, which were
later assigned the designations of TWA~26 and 27, respectively \citep{mam05}.
During a survey for substellar companions with adaptive optics imaging,
\citet{cha04} discovered a faint candidate with a projected separation of 
$0\farcs78$ from TWA~27. They verified its cool nature through near-infrared 
(IR) spectroscopy and estimated a mass of $\sim5$~$M_{\rm Jup}$ from
photometry, which would make it the first directly imaged planetary mass
companion outside of the solar system.
Companionship was confirmed through additional epochs of imaging that
demonstrated a common proper motion with the primary \citep{cha05,son06}.
Although the secondary is well within the mass range of extrasolar planets 
\citep{how10}, it likely formed in the manner of a binary star rather than
as a planet in a circumstellar disk given the large mass ratio of the pair
($\sim0.2$) compared to the typical value of 
$M_{\rm disk}/M_{\rm star}$ for low-mass stars and brown dwarfs 
\citep[$\sim0.01$,][]{and13,moh13}, as discussed in \citet{lod05}.
The mass ratio of the TWA~27 system is also well above the 
upper limit of $\sim$0.04 in the International Astronomical Union's working 
definition of an exoplanet \citep{lec22}.

Since its discovery, TWA~27B has been the subject of numerous studies to
characterize its stellar and atmospheric properties. 
One parameter that is important in such analysis is the distance,
which has been estimated for the primary using the moving cluster method
\citep{mam05,mam07} and parallax \citep{bil07,giz07,duc08}.
Currently, the most accurate distance for TWA~27A is 64.5$\pm0.4$~pc
\citep{bai21}, which is based on a parallax measurement from the
third data release of the Gaia mission \citep{gaia16b,bro21,val22}.

For all of the distances previously adopted for TWA~27B, which have
ranged from $\sim$53--70~pc, the luminosity derived from its near-IR photometry 
and a bolometric correction for field dwarfs has
been much lower than predicted for an object at the age of TWA and at 
the effective temperature implied by both model fits to its spectrum
\citep{moh07,pat12} and the combination of its spectral type 
\citep[mid-to-late L,][]{cha04,moh07,pat10,all13} and the temperature 
scale for field dwarfs \citep{gol04}. Initial attempts were made to
explain this seemingly anomalous property \citep{mam07,moh07}. 
However, the discovery of additional young L/T dwarfs and planets has
demonstrated that TWA~27B is not unusual for its mass and age
\citep{met06,luh07,met09,bow10,cur11,ske11,ske14,fah12,fah16,liu13,liu16,fil15}.
The discrepancies between the inferred temperatures and luminosities of
young late-type objects like TWA~27B and the values predicted by evolutionary
models likely reflect age dependencies of the temperature scale and
the near-IR bolometric corrections. Namely, at younger ages, a given
spectral type corresponds to a cooler temperature and a smaller fraction
of the flux is emitted at near-IR wavelengths.
The spectral energy distributions (SEDs) of young late-type objects have been
broadly reproduced by models of cloudy, low-gravity atmospheres that 
experience non-equilibrium chemistry 
\citep{bar11a,bar11b,mad11,mar12,ske11,ske14,cha18}
as well as models of cloudless atmospheres that have a temperature
gradient reduction caused by fingering convection \citep{tre17}.

TWA~27B continues to play an important role in understanding the early
evolution of planetary mass objects because of its well-determined distance
and age and the paucity of known objects with a similar combination of 
age and mass. In this paper, we seek to use the Near-Infrared Spectrograph 
\citep[NIRSpec,][]{jak22} on board the James Webb Space Telescope 
\citep[JWST,][]{gar23} to measure the SED of TWA~27B more accurately 
and across a wider range of wavelengths than done in previous studies.

\section{Observations}

The TWA~27 system was observed with NIRSpec's integral-field unit 
\citep[IFU,][]{bok22} through guaranteed time observation program 1270
(PI: S. Birkmann) on 2023 February 7 (UT).
The IFU has a large enough field of view ($3\farcs1\times3\farcs2$) that
it was able to observe both components of the system at the same time.
Data were collected with the following grating/filter combinations,
which provided a spectral resolution of $\sim$2700:
G140H/F100LP (0.97--1.89~\micron), G235H/F170LP (1.66--3.17~\micron),
and G395H/F290LP (2.87--5.27~\micron). 
The resulting data have gaps at 1.42--1.47, 2.39--2.47, and 4.03--4.17~\micron\ 
from the physical separation between the two detector arrays.
For each configuration, we used either 31 or 32 groups, four dithers, and 
the NRSIRS2RAPID readout pattern, resulting in a total exposure time
of $\sim$1900~s. To facilitate the subtraction of TWA~27A
when extracting the spectrum of its companion, identical observations
were performed for TWA~28 \citep[SSSPMJ1102$-$3431,][]{sch05a}, which
has a similar spectral type and brightness as TWA~27A.
The position angle of the telescope was selected to avoid bright stars
falling within the field of view of the microshutter assembly.

\section{Data Reduction}
\label{sec:reduction}

We have reduced the NIRSpec data for the TWA sources as well as similar
observations of the planetary mass companion VHS J125601.92$-$125723.9b
\citep[VHS~1256b,][]{gau15} from program 1386 \citep{hin22}, which
were first reported by \citet{mil23}. It is useful to compare the spectra
of TWA~27B and VHS~1256b since they have roughly similar spectral types but
different ages.

\begin{figure}
\epsscale{0.6}
\plotone{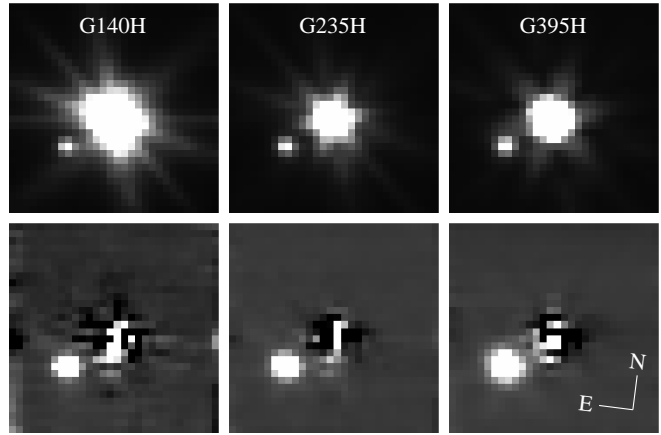}
\caption{NIRSpec IFU images of TWA~27A and B after collapsing along the
dispersion direction, before and after PSF subtraction of the primary
(top and bottom). The size of each image is $3\arcsec\times3\arcsec$.}
\label{fig:cube}
\end{figure}

For the data reduction, we began by retrieving the {\tt uncal} files from the
Mikulksi Archive for Space Telescopes (MAST):
\dataset[doi:10.17909/jzze-5611]{http://dx.doi.org/10.17909/jzze-5611}.
We performed ramps-to-slopes 
processing on those files using the NIRSpec Instrument Pipeline
Software{\footnote{\url{https://jwst-tools.cosmos.esa.int/index.html}},
which features a correction for ``snowballs" \citep{bok23}} 
and a correction for residual correlated noise in IFU data. 
The resulting count rate maps ({\tt rate} files) were used as inputs to the 
JWST {\tt calwebb\_spec2} 
pipeline with the default processing steps, including assignment of the World
Coordinate System, flat fielding, and aperture correction. 
We used the JWST Science Calibration pipeline version 1.9.4 under Calibration
Reference Data System context jwst\_1063.pmap. We flagged outliers 
in the resulting {\tt cal} files and then constructed the IFU data cubes using
the {\tt calwebb\_spec3} pipeline with outlier detection disabled. 
The latter was necessary as currently the pipeline's outlier detection is
not working properly, resulting in many erroneously flagged data points
and poor final cubes. We selected a spaxel size of $0\farcs1$ for the
cubes and used the ``ifualign" coordinate system to provide the same point 
spread function (PSF) orientation among observations of different targets. 
Spectra for TWA~27A, TWA~28, and VHS~1256b were extracted from these
cubes with an aperture radius of 4 spaxels. 
Before extracting TWA~27B, PSF subtraction was applied to the primary 
using a version of the data cube for TWA~28 that had been scaled and shifted
appropriately. In Figure~\ref{fig:cube}, we show the 2-D images of 
TWA~27A and B for each of the three gratings before and after PSF subtraction.
Flux calibration was performed with observation 62 for 
standard star P330E from program 1538 (PI: Karl Gordon), which utilized 
the NIRpec IFU with the same gratings and filters as our observations.
The reduced spectra of TWA~27B and VHS~1256b are available at
\dataset[doi:10.5281/zenodo.7876159]{https://doi.org/10.5281/zenodo.7876159}.

We have compared the reduced spectra of VHS~1256b from this work and 
\citet{mil23}. For some wavelength ranges, our version of the spectrum
has less low-order structure and exhibits a more smoothly varying continuum.
For instance, the absorption feature near 1.66~\micron\ that was noted by
\citet{mil23} is absent from our reduction.

We have derived synthetic photometry from our reduced spectra of
the TWA sources and VHS~1256b, which we use to assess the accuracy
of the absolute fluxes in these data. Such measurements also facilitate
comparisons to other late-type objects (Section~\ref{sec:compare2}).
We have selected the following bands for synthetic photometry:
$JHK_s$ from 2MASS, $JHK$ from the Mauna Kea Observatories (MKO) system 
\citep{tok02}, the W1 and W2 bands (3.4 and 4.6~\micron) for the Wide-field
Infrared Survey Explorer \citep[WISE,][]{wri10} and the reactivated mission 
\citep[NEOWISE,][]{mai14}, and most of the medium- and wide-band filters for 
NIRCam on JWST \citep{rie05,rie23}. 2MASS, MKO, and WISE are the most 
frequently measured filters for substellar objects while the NIRCam
photometry may be useful for comparison to data from that camera in future 
studies. When calculating the photometry, we interpolated across the
three gaps in the spectra.

\begin{figure}
\epsscale{1.2}
\plotone{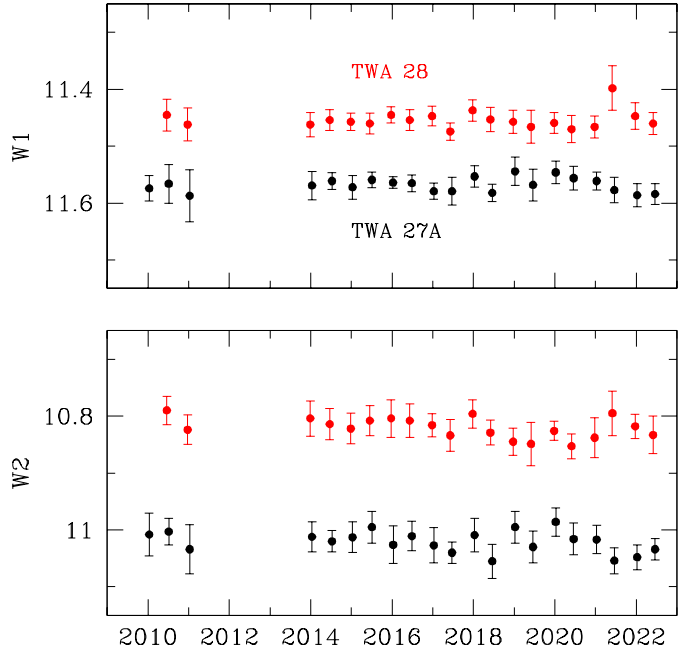}
\caption{Medians and standard deviations of WISE/NEOWISE data for
TWA~27A and TWA~28 among the single exposures taken during epochs spanning
$\sim1$--4~days.
}
\label{fig:var}
\end{figure}

For comparison to our synthetic photometry, we have considered $JHK_s$
data for TWA~27A and TWA~28 from the 2MASS Point Source Catalog \citep{skr03},
$JHK$ data for VHS~1256b from the sixth data release of the
Visible and Infrared Survey Telescope for Astronomy (VISTA) Hemisphere 
Survey \citep{mcm13}, and the mean 4.5~\micron\ photometry
for VHS~1256b from \citet{zho20} after conversion to W2 \citep{kir21}.
The mean photometry from \citet{zho20} was based on continuous measurements 
during a 36~hr period. WISE and NEOWISE have provided photometry in W1 and W2
for TWA~27A and TWA~28 at multiple epochs between 2010 and 2022. 
The data are clustered in epochs that span 1--4~days, each containing 11--31 
single exposures. TWA~27A and TWA~28 were observed at 21 and 20 epochs,
respectively. For each epoch and band, we have calculated the median and 
standard deviation of the single-exposure measurements from the AllWISE 
Multiepoch Photometry Table \citep{wri13}
and the NEOWISE-R Single Exposure Source Table \citep{cut23}.
The resulting measurements are plotted versus time in Figure~\ref{fig:var}.
For individual epochs, the standard deviations are typically 0.02--0.04~mag,
which serve as upper limits on the variability on timescales of hours.
For each object and band, the standard deviation of the medians across
all epochs is $\lesssim0.02$~mag. 
The medians of the medians across all epochs are W1=11.58 and W2=11.03 for
TWA~27A and W1=11.46 and W2=10.83 for TWA~28, which are adopted for
comparison to the NIRSpec photometry.

The differences between the synthetic photometry from NIRSpec and the 
photometry from previous measurements, $m_{\rm NIRSpec}-m_{\rm phot}$, are
0.05 ($J$), 0.04 ($H$), 0.06 ($K_s$), $-$0.03 (W1), and 0.00 (W2) for TWA~27A,
0.02 ($J$), 0.03 ($H$), 0.04 ($K_s$), $-$0.07 (W1), and $-$0.04 (W2) for TWA~28,
and 0.09 ($J$), $-$0.11 ($H$), $-$0.05 ($K$), and 0.05 (W2) for VHS~1256b.
These results do not show a systematic offset across all bands and objects, 
indicating that the flux calibrations of the spectra are fairly accurate. 
The differences in photometry can be plausibly explained by variability.
In particular, the wide range of offsets for VHS~1256b is not surprising
given that it is known to exhibit
large wavelength-dependent variability \citep{bow20,zho20,zho22a}.
The synthetic photometry in Vega magnitudes from the NIRSpec data
is presented in Table~\ref{tab:phot}.
The photometric errors should be dominated by the uncertainty in the flux
calibration for the NIRSpec data, for which we have adopted a value of 3\%.

\section{Analysis}

\subsection{Comparison to Previous Spectroscopy}
\label{sec:compare1}

\begin{figure}
\epsscale{1.2}
\plotone{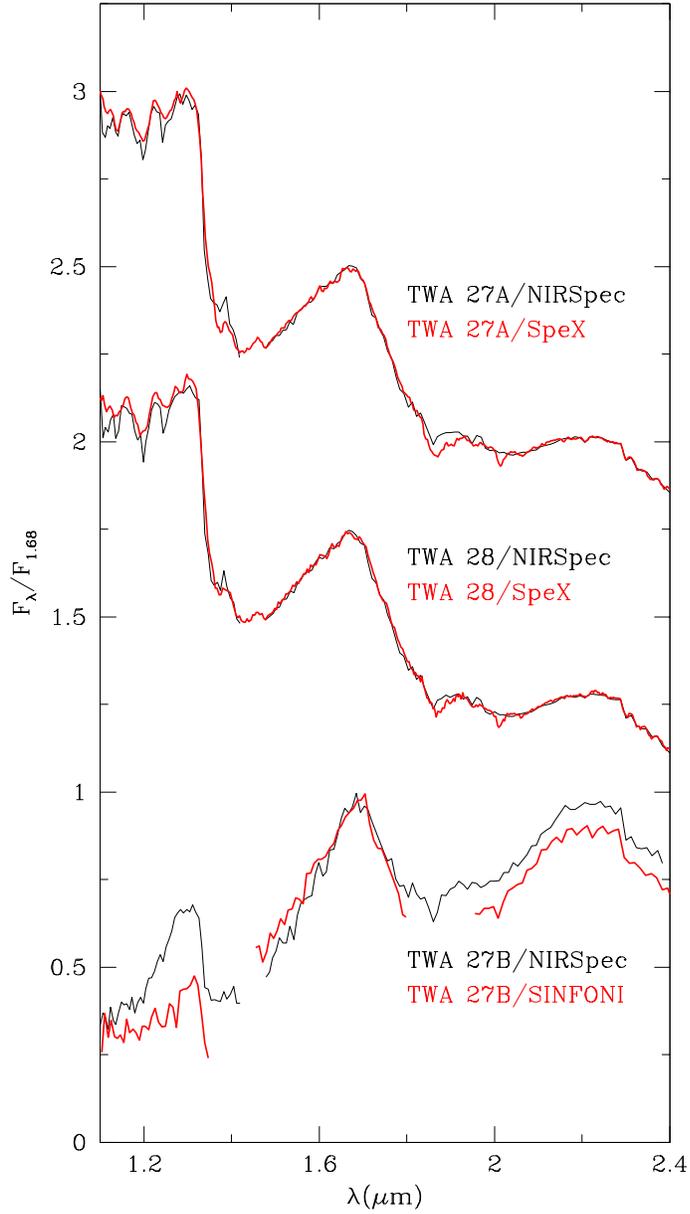}
\caption{Comparison of near-IR spectra from NIRSpec (this work)
and ground-based instruments \citep{loo07,pat10,luh17}.
The NIRSpec and SINFONI data have been binned to a resolution of R=200.
The relative flux calibrations of the $J$ and $HK$ spectra from SINFONI were
based on uncertain photometry in $J$ and $K$.}
\label{fig:spec1}
\end{figure}

Extensive spectroscopy has been performed on TWA~27A and TWA~28
from UV through mid-IR wavelengths 
\citep{giz02,sch05b,loo07,mor08,her09,ric10,bon14,her14,ven19}
while more limited spectra have been collected at 1-2.5~\micron\ for TWA~27B 
\citep{cha04,moh07,pat10}.
To compare the broad near-IR spectral shapes measured by NIRSpec and
other spectrographs, we show in Figure~\ref{fig:spec1} the
NIRSpec data and low-resolution spectra from SpeX \citep{ray03} at the NASA
Infrared Telescope Facility for TWA~27A and TWA~28 \citep{loo07,luh17} and a 
spectrum from SINFONI \citep{eis03} on the Very Large Telescope for
TWA~27B \citep{pat10}. The NIRSpec and SINFONI spectra have been binned 
to a resolution of R=200. The NIRSpec and SpeX data agree very closely.
The SINFONI spectrum for TWA~27B is significantly redder than the NIRSpec
data from 1.2--1.7~\micron, which is likely caused by the large uncertainties
in the $J$ and $K$ photometric measurements \citep{cha04,moh07}
that were used to determine the relative flux calibration of the $J$ 
and $HK$ spectra from SINFONI. Meanwhile, the NIRSpec data are somewhat
redder than the SINFONI spectrum at 1.7--2.4~\micron.
TWA~27B had $J-K=3.07\pm0.23$
based on the photometry from \citet{cha04} and \citet{moh07},
which has made it one of the reddest known L-type objects.
However, we derive $J-K=2.72$ (MKO) from the NIRSpec data,
indicating that TWA~27B is not quite as red as previously reported.

\subsection{Comparison to Young Late-type Objects}
\label{sec:compare2}

\begin{figure}
\epsscale{1.2}
\plotone{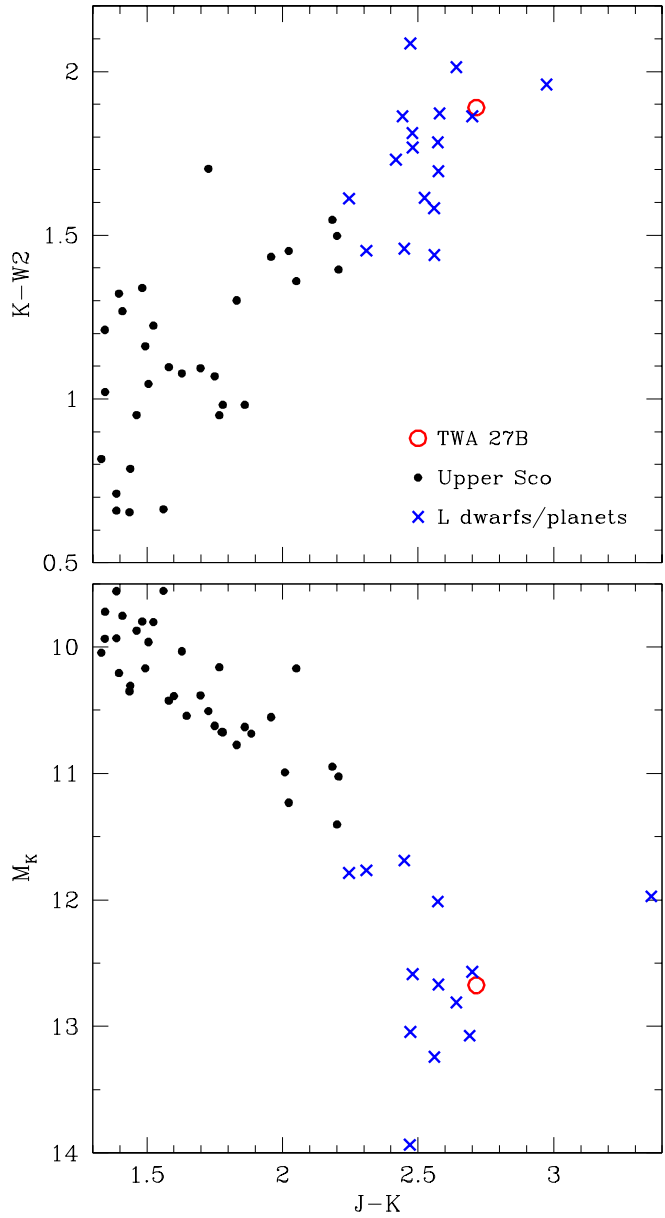}
\caption{
$K-$W2 and $M_K$ versus $J-K$ for TWA~27B, members of
Upper Sco \citep[][K. Luhman, in preparation]{luh22sp}, and red L-type
dwarfs and planets \citep[][references therein]{sch23}.
Some of the L-type objects appear in only one diagram because they lack
parallax measurements or W2 photometry.}
\label{fig:cmd}
\end{figure}

The photometric properties of young late-type objects can be examined with
IR color-color and color-magnitude diagrams (CMDs). In Figure~\ref{fig:cmd}, 
we have plotted TWA~27B on diagrams of $K-$W2 and $M_K$ versus $J-K$ using the
synthetic photometry measured with NIRSpec. The $J$ and $K$ data are on the
MKO system. For comparison, we have included the sample of L-type dwarfs and
planetary mass companions with $J-K>2.2$ compiled by \citet{sch23} (most have
ages of $\lesssim200$~Myr) and the known members of Upper Sco 
\citep[$\gtrsim$M9,][K. Luhman, in preparation]{luh22sp} that are located 
within the central triangular field of the association defined by 
\citet{luh20u}. Upper Sco offers the largest available sample of planetary mass 
brown dwarfs near the age of TWA. When calculating $M_K$ for Upper Sco members,
we have used the parallactic distances from \citet{bai21} when available
and otherwise have adopted the median distance of 144~pc for the members
of Upper Sco within the field in question. Among the L-type objects compiled
by \citet{sch23}, some of the free-floating dwarfs lack parallax measurements 
(absent from the CMD) and some of the companions lack photometry
in W2 (absent from $K-$W2 versus $J-K$).

Most of the objects in Figure~\ref{fig:cmd} form a fairly well-defined locus
in each diagram. In the CMD, the locus begins to turn back
to bluer colors with the faintest object, HD~8799b, which is the behavior
observed in the transition from L to T types among normal field dwarfs
\citep{bur02}. The red outlier in $J-K$ is HD~206893B \citep{mil17}. 
TWA~27B appears along the red edge of the locus and is less extreme in $J-K$ 
using our photometry than in previous studies, as mentioned in 
Section~\ref{sec:compare1}.  
The data in Figure~\ref{fig:cmd} illustrate that TWA~27B is not anomalous 
relative to other young planetary mass objects, which has become widely 
recognized over the last decade.

\begin{figure}
\epsscale{1.2}
\plotone{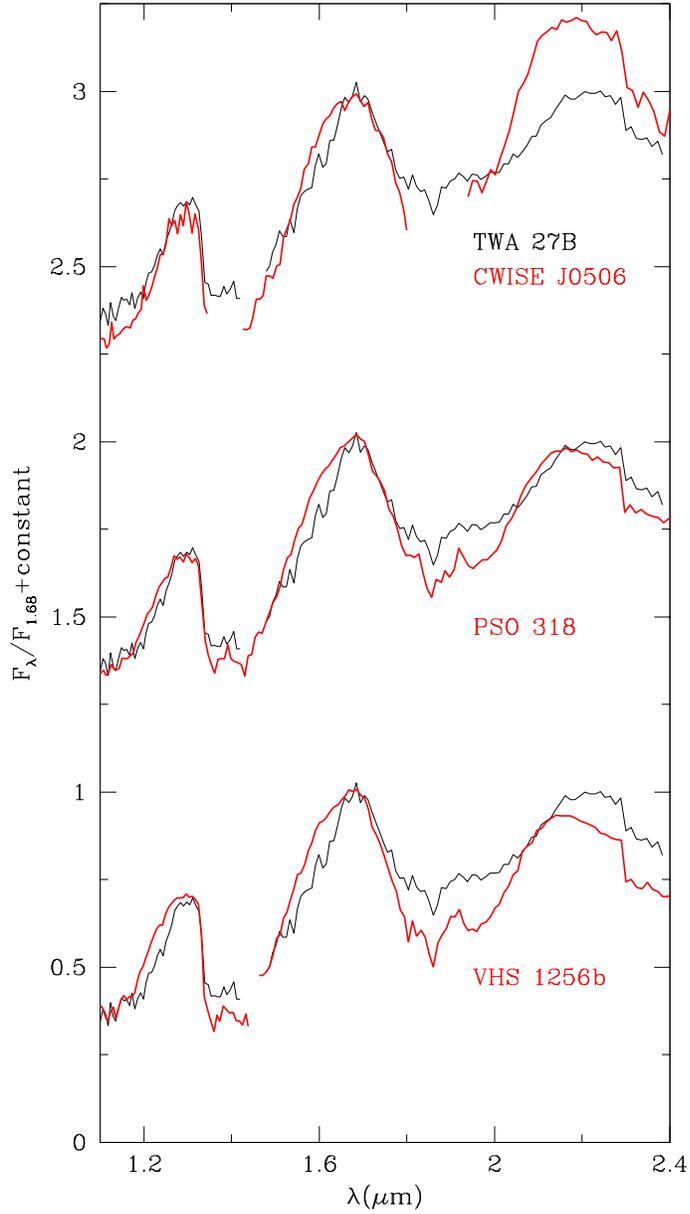}
\caption{Near-IR spectrum of TWA~27B from JWST/NIRSpec (this work)
compared to spectra of CWISE J0506 \citep{sch23}, PSO~318 \citep{liu13}, 
and VHS~1256b (this work; see also \citet{mil23}).
The spectra have been binned to a resolution of R=200.}
\label{fig:spec2}
\end{figure}

In Figure~\ref{fig:spec2}, we compare the near-IR spectrum of TWA~27B
to data for two of the reddest L dwarfs from Figure~\ref{fig:cmd},
CWISE J050626.96+073842.4 \citep[CWISE 0506,][]{sch23} and 
PSO J318.5338$-$22.8603 \citep[POS 318,][]{liu13}.
We also include our reduction of the NIRSpec observations of VHS~1256b since
it is the only object from Figure~\ref{fig:cmd} other than TWA~27B that
has NIRSpec data available. Since we are comparing their broad shapes,
the spectra have been binned to a resolution of R=200.
TWA~27B is bluer than CWISE 0506 and slightly redder than PSO~318
and VHS~1256b, which is consistent with their $J-K$ colors. 
The shapes of the $H$- and $K$-band continua differ somewhat
between TWA~27B and the other objects, which is likely because of older
ages for the latter. Both PSO~318 and CWISE~0506 have been classified as
likely members of the $\beta$~Pic association \citep{all16,sch23}, 
which has an age of $\sim$21--24~Myr \citep{bin16}, while an age of
$140\pm20$~Myr has been estimated for the VHS~1256 system \citep{dup23}

\begin{figure}
\epsscale{1.2}
\plotone{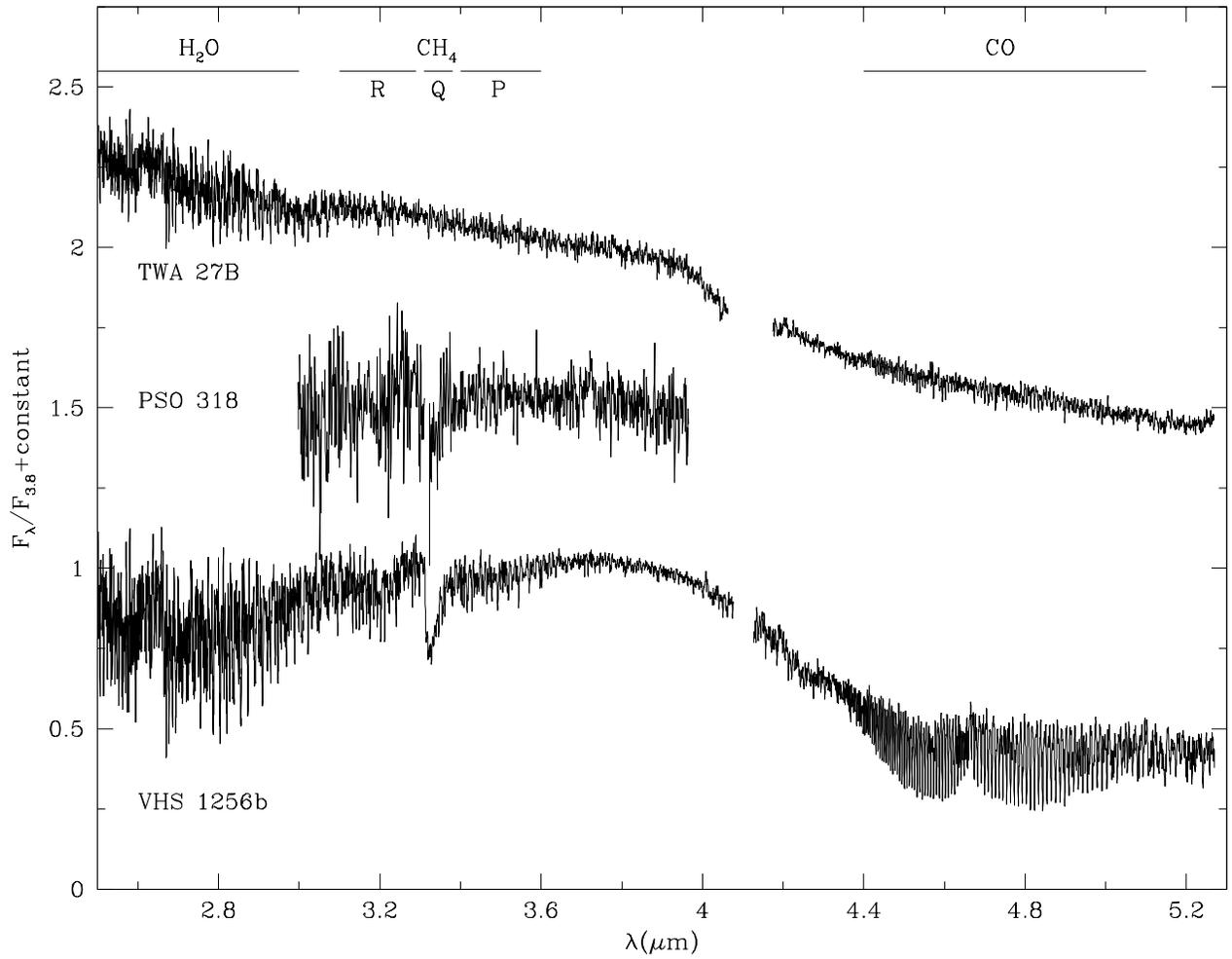}
\caption{Mid-IR spectra of TWA~27B (this work), PSO~318 \citep{bei23},
and VHS~1256b (this work; see also \citet{mil23}).}
\label{fig:spec3}
\end{figure}

Spectroscopy has not been performed previously on TWA~27B longward of the $K$
band. In Figure~\ref{fig:spec3}, we show its 2.5--5.3~\micron\ spectrum from
NIRSpec with the NIRSpec data for VHS~1256b and a 3--4~\micron\ ground-based 
spectrum of PSO~318 \citep{bei23} (see also \citet{mil18}).
CWISE~0506 has not been observed spectroscopically at these wavelengths.
Although TWA~27B and VHS~1256b have roughly similar near-IR spectra, their
3--5~\micron\ spectra exhibit notable differences.
VHS~1256b has strong absorption in the $Q$-branch of the $\nu_3$ fundamental
band of CH$_4$ at 3.33~\micron, broader shallow absorption in the
surrounding $P$- and $R$-branches, and strong absorption from the fundamental
band of CO at 4.4--5.2~\micron, as discussed by \citet{mil23} in their analysis 
of the NIRSpec data.
In TWA~27B, CH$_4$ absorption is not detected and the CO lines are much weaker.
The CH$_4$ absorption in PSO~318 appears to be similar to or slightly
weaker than that of VHS~1256b. The origins of these differences in CH$_4$
and CO are discussed in the next section.

\subsection{Comparison to Model Spectra}
\label{sec:model}

As mentioned in Section~\ref{sec:intro}, it has been possible to identify
both cloudy and cloudless models that can reproduce the general shapes of
the SEDs of young L-type objects. In the case of TWA~27B, modeling has been
applied to the near-IR spectra from \citet{pat10} and \citet{moh07}
and the 3--4~\micron\ photometry from \citet{cha04} and \citet{ske14}
\citep{moh07,bar11b,ske11,ske14,pat10,pat12}. In this study, we have attempted 
to fit the 1--5~\micron\ data for TWA~27B from NIRSpec with spectra predicted 
by the {\tt ATMO} models of cloudless brown dwarfs \citep{tre15,tre17}.
In these models, diabatic convective processes \citep{tre19} induced by
non-equilibrium chemistry of CO/CH$_4$ and N$_2$/NH$_3$ reduce the
temperature gradient in an atmosphere and reproduce the red near-IR
colors of L dwarfs \citep{tre16}. \citet{pet23} described a recent grid of
the models in which the parameters were effective temperature 
($T_{\rm eff}$), surface gravity (log~$g$), effective adiabatic index 
($\gamma$), eddy diffusion coefficient ($K_{\rm zz}$), metallicity ([M/H]),
and C/O ratio. We utilized that model grid for an initial fit to
the NIRSpec data for TWA~27B and we calculated additional models to
further refine the fit, converging on a model with $T_{\rm eff}=1300$~K, 
log~$g=3.5$, $\gamma=1.03$, $K_{\rm zz}=10^5$, [M/H]=0.2, and a solar value
for C/O. The absolute fluxes of the NIRSpec data lead to a radius of 
0.116~$R_\odot$ for that model.  

\begin{figure}
\epsscale{1.2}
\plotone{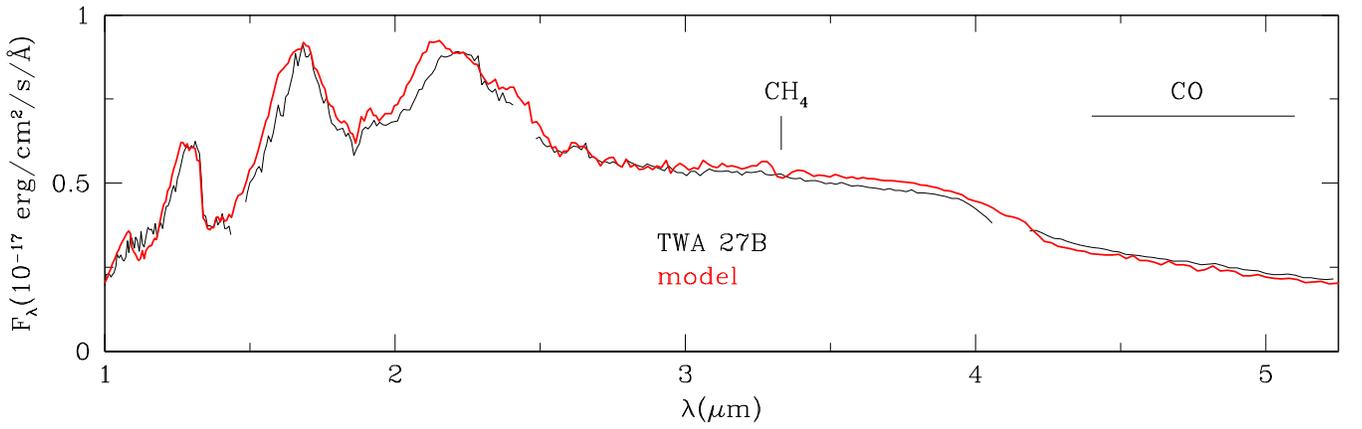}
\caption{IR spectrum of TWA~27B from JWST/NIRSpec compared to a model
spectrum for a cloudless atmosphere with $T_{\rm eff}=1300$~K, log~$g=3.5$, 
$\gamma=1.03$, $K_{\rm zz}=10^5$, [M/H]=0.2, a solar value for C/O, 
and $R=0.116$~$R_\odot$ \citep{tre17,pet23}. 
The spectra have been binned to a resolution of R=200.}
\label{fig:spec5}
\end{figure}

In Figure~\ref{fig:spec5}, we show the NIRSpec and model spectra after
binning both to a resolution of R=200.
The model agrees quite well with the data overall. 
The primary differences consist of the shapes of the $H$- and $K$-band
continua, the flux at 3--4~\micron, and the strengths of CH$_4$
at 3.33~\micron\ and CO at 4.4--5.2~\micron, where both of the latter
features are stronger in the model.
The inclusion of clouds could potentially improve the fit since
they should reduce the strengths of those features \citep{bar11b}.
The temperature and radius of the model for TWA~27B correspond to
an age of $\sim$100~Myr according to evolutionary models \citep{cha23}, 
which is much older than the age of $\sim$10~Myr implied by a comparison of
the primary to evolutionary models and various age constraints
for the association \citep[e.g.,][]{luh23twa}.
This discrepancy likely reflects remaining deficiencies in the adopted
atmospheric model such that the true temperature and radius of TWA~27B are
lower and higher, respectively, than the values implied by our SED fitting. 
For instance, if we assume an age of 10 Myr,
our estimate of the luminosity of TWA~27B in Section~\ref{sec:mass}
implies a temperature of 1200~K according to evolutionary models, 
whereas 1300~K was derived from our model fit to the NIRSpec data.
Early modeling of TWA~27B's SED exhibited a larger discrepancy of this kind
\citep[$T_{\rm eff}\sim1600$~K,][]{moh07,pat10}
while models that included non-equilibrium chemistry and thick clouds 
produced lower temperatures that were less discrepant 
\citep[$T_{\rm eff}\sim1000$~K,][]{bar11b}.

\citet{mil18,mil23} found that the 3.33~\micron\ CH$_4$ band is weaker in 
VHS~1256b than in field L dwarfs, which they attributed to non-equilibrium
chemistry in VHS~1256b. Weak CH$_4$ also has been observed
in other young L dwarfs with ages of $\sim$10--100~Myr \citep{bei23}.
That trend of weaker CH$_4$ with younger age continues with TWA~27B, 
where the feature is absent.
In addition to weakening CH$_4$, an enhancement of
non-equilibrium chemistry should strengthen the fundamental band of CO
\citep{ske14,muk22}, so the fact that both CH$_4$ and CO are significantly
weaker in TWA~27B than in VHS~1256b indicates the influence of
an additional atmospheric property that varies with age, such
as the temperature gradient \citep{tre15,tre17} or cloud
thickness \citep{bar11b,ske14,cha18}.

\subsection{Evidence of Circumstellar Disks}

Spectra like those collected with NIRSpec in TWA can contain
signatures of circumstellar disks in the form of emission lines or
mid-IR emission in excess above that expected from a stellar photosphere.
TWA~27A and TWA~28 have previous evidence of disks
from some combination of UV emission lines and continuum
\citep{giz05,her09,fra10,ven19}, H$\alpha$ emission \citep{moh03,sch05b,ste07}, 
mid- and far-IR excess emission \citep{ste04,ria06b,ria08,mor08,har12}, 
and millimeter emission \citep{moh13,ric17}.
An outflow also has been detected from TWA~27A \citep{whe07,whe12}.
The near-to-mid-IR colors derived from the NIRSpec data for
TWA~27A and TWA~28 (e.g., $J-$W1 and $J-$W2) are similar to colors from
2MASS, WISE, and the Spitzer Space Telescope, 
and thus exhibit color excesses from disks that are consistent with
those measured in previous studies \citep{ria06b,sch12a}.

Given its faintness and small separation from the primary, TWA~27B has
had few constraints on the presence of a disk.
\citet{ric17} performed sensitive observations with
the Atacama Large Millimeter Array (ALMA) that were capable of resolving 
the components of TWA~27, but no emission was detected from the secondary.
They derived a 3~$\sigma$ upper limit of $\sim0.013$~$M_\Earth$
($\sim1$~$M_{\rm Moon}$) on the mass of dust surrounding TWA~27B.
By extending longward of previous IR measurements, the NIRSpec data
at 4--5~\micron\ provide a new constraint on disk emission. 
Our synthetic NIRSpec photometry in the W2 band spans that wavelength range. 
The color-color diagram in Figure~\ref{fig:cmd}
shows that TWA~27B does not exhibit a color excess in $K-$W2 relative
to other young objects with similar values of $J-K$, which indicates
that it lacks excess emission in W2. 

\begin{figure}
\epsscale{1.4}
\plotone{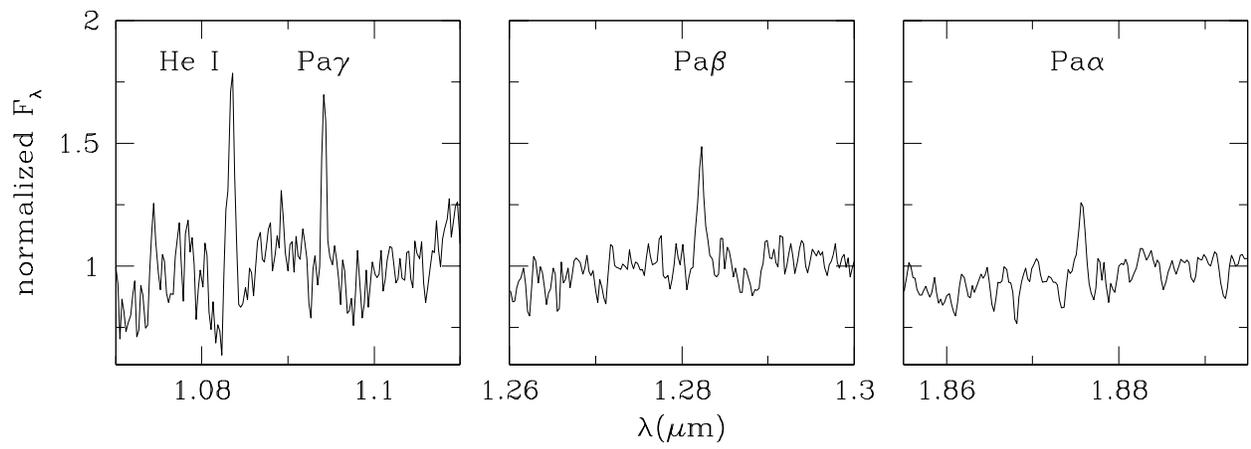}
\caption{Detections of He~I and Paschen emission lines in the
JWST/NIRSpec spectrum of TWA~27B.}
\label{fig:spec4}
\end{figure}

Although the available mid-IR and millimeter data for TWA~27B have not detected
disk emission, the NIRSpec spectrum does contain detections of emission
in Paschen transitions ($\alpha$, $\beta$, $\gamma$, and possibly $\delta$) 
and the (blended) He~I triplet at 1.083~\micron, as shown in 
Figure~\ref{fig:spec4}. These emission lines are signatures of accretion 
when observed in young stars \citep{nat04,edw06}.
Estimating an accretion rate from the strength of the He~I emission is not
straightforward because it is affected by both accretion and winds
\citep{edw06,erk22,tha22}. Relations between Pa$\beta$ line luminosity and
accretion luminosity are available for stars and brown dwarfs
\citep{nat04,alc17}.
It is unknown whether such relations are applicable to brown dwarfs as small 
in mass as TWA~27B, but we have used them to derive an accretion luminosity
from its Pa$\beta$ line flux ($6.6\pm1.2\times10^{-17}$ erg~cm$^{-2}$~s$^{-1}$) 
and the distance of TWA~27A. We have converted the accretion luminosity 
to an accretion rate in the manner done in \citet{gul98} by assuming a
mass of 0.005~$M_\odot$ and a radius of 0.14~$R_\odot$ for TWA~27B, which
are expected from evolutionary models for the luminosity and age of the
object (Section~\ref{sec:mass}). The resulting accretion rate ranges from 
$\dot{M}\sim10^{-13}$--$10^{-12}$~$M_\odot$~yr$^{-1}$, which appears
to be plausible when extrapolating the relationship between 
stellar mass and accretion rate for young stars and brown dwarfs to
the mass of TWA~27B \citep{muz05,her08,har16}. 
Given the dust mass upper limit from ALMA and
a standard gas-to-dust ratio of $\sim$100, a disk around TWA~27B could
undergo accretion at these rates for $\lesssim$4--40~Myr, which is
comparable to the age of TWA. Thus, steady accretion at the rate
implied by Pa$\beta$ appears plausible given the mass constraints from ALMA.
If TWA~27B does have a disk, it likely exhibits excess emission in unpublished
5--28~\micron\ data that were obtained with the Mid-infrared Instrument on 
JWST \citep[MIRI,][]{rie15} during the same visit as the NIRSpec observations.

TWA~27B joins a small but growing sample of young planetary mass companions 
in which possible evidence of accretion has been detected via emission lines
\citep{bow11,lac15,wag18,haf19,eri20,chi21,cur22,bet22,zho14,zho21,zho22b,rin23}.
As an aside, we note that one of those companions, 
2MASS J01033563$-$5515561 C \citep{del13}, appears to lack a spectral
classification but does have optical spectroscopy available
\citep{eri20}. We have measured a spectral type of L0$\pm0.5$ from those
data based on a comparison to young L-type standards \citep{cru09,cru18}.

\subsection{Luminosity and Mass Estimates for TWA~27B}
\label{sec:mass}

Previous studies have estimated the luminosity of TWA~27B by
combining a near-IR band (typically $K$) with a bolometric correction for
L dwarfs and an adopted distance.  They then derived a mass by
comparing the luminosity to the values predicted by evolutionary models for
the age of TWA \citep{cha04,mam05,mam07,duc08,bar11b}.
The resulting mass estimates have spanned from $\sim$3--7~$M_{\rm Jup}$.
Since the NIRSpec data have measured the SED of TWA~27B across a wide
range of wavelengths and the primary has a precise distance measurement
from Gaia, we can improve the estimates of the luminosity and mass.
We have calculated the luminosity of TWA~27B by integrating the flux
in the NIRSpec spectrum (1--5.3~\micron) and the best fitting model
spectra at shorter and longer wavelengths and applying the distance
from \citet{bai21}, arriving at log~$L/L_\odot=-4.466\pm0.014$.
We have adopted the distance error from \citet{bai21} and a 3\% uncertainty
in the flux calibration of the NIRSpec data.
Our luminosity estimate is similar to those from previous studies after
accounting for the differences in adopted distances.

\begin{figure}
\epsscale{1.2}
\plotone{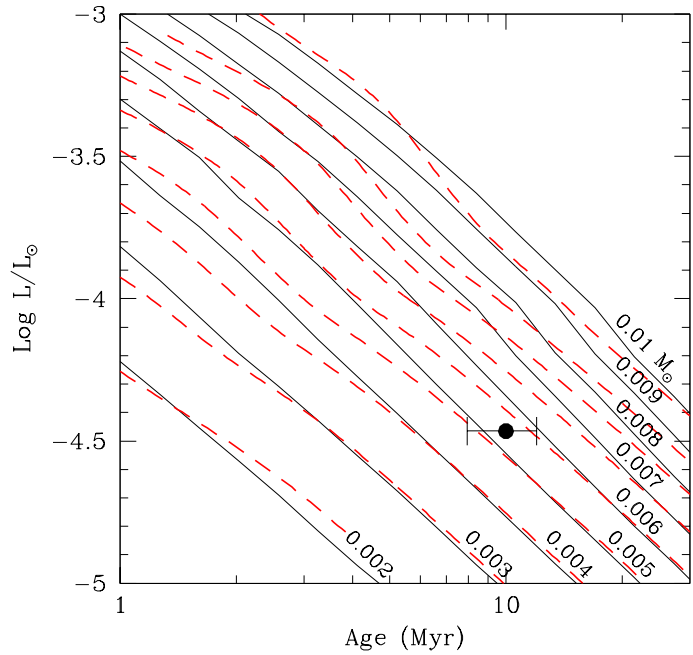}
\caption{Luminosity estimate for TWA~27B compared to luminosities
as a function of age predicted by evolutionary models of
\citet{cha23} (solid lines) and \citet{bur97} (dashed lines).}
\label{fig:mass}
\end{figure}

For TWA~27B, we have adopted an age of 10$\pm$2~Myr that has been
estimated for its association \citep{luh23twa}.
In Figure~\ref{fig:mass}, we have plotted TWA~27B with the luminosities 
predicted as a function of age for masses of 0.002--0.01~$M_\odot$
by the evolutionary models of \citet{cha23} and \citet{bur97}.
Both sets of models imply a mass of 0.005--0.006~$M_\odot$ (5--6~$M_{\rm Jup}$) 
for TWA~27B. For an age of 10~Myr, our luminosity estimate
corresponds to a temperature of 1200~K according to the evolutionary models,
which is lower than the value of 1300~K derived from our model fitting of the
SED, as mentioned in Section~\ref{sec:model}.

\section{Conclusions}

We have used NIRSpec on JWST to perform 1--5~\micron\ spectroscopy
on the young planetary mass companion TWA~27B.
We also have reduced similar data for VHS~1256b \citep{mil23},
which has a roughly similar spectral type to TWA~27B but an older age
($\sim140$~Myr). Our results are summarized as follows:

\begin{enumerate}

\item
We have calculated synthetic photometry from the NIRSpec spectra for 
TWA~27A, TWA~27B, TWA~28, and VHS~1256b in a variety of bands between 
1--5~\micron. Given the large errors and limited wavelength range
of its previous photometry, the new measurements for TWA~27B are particularly
useful for comparison to other young planetary mass objects.
We find that TWA~27B is not as red in $J-K$ as reported in previous studies,
making it less extreme in that color relative to other young L dwarfs. 
In diagrams of $K-$W2 and $M_K$ versus $J-K$, it falls within the 
sequence formed by young L-type dwarfs and planets.

\item
NIRSpec has provided the first spectroscopy of TWA~27B at 2.5--5~\micron.
In these data, the fundamental band of CH$_4$ is absent and the fundamental 
band of CO is weak, whereas both features are stronger in VHS~1256b.
The nondetection of CH$_4$ reinforces a previously observed trend of weaker
CH$_4$ with younger ages among L dwarfs \citep{mil18,mil23,bei23}, which has 
been attributed to enhanced non-equilibrium chemistry among young objects.  
Explaining the weak CO in TWA~27B requires that an additional atmospheric 
property varies with age, such as the temperature gradient or cloud thickness.

\item
We have compared the NIRSpec data for TWA~27B to spectra predicted by the 
{\tt ATMO} models of cloudless brown dwarfs \citep{tre15,tre17},
which include non-equilibrium chemistry and a temperature gradient
reduction caused by fingering convection. The broad shape of the spectrum 
is matched well with a model that has $T_{\rm eff}=1300$~K and a low surface 
gravity. One of the primary differences between the model and observed spectra
lies in the fundamental bands of CH$_4$ and CO, which are somewhat stronger in 
the model spectrum. Meanwhile, the model temperature of 1300~K is higher than
expected from evolutionary models given the luminosity and age of TWA~27B
($T_{\rm eff}=1200$~K).
Based on previous modeling of young L-type objects like TWA~27B, the inclusion
of thick clouds could potentially resolve both issues \citep[e.g.,][]{bar11b}.
It remains to be seen whether cloudy models can provide a good fit to the
1--5~\micron\ data from NIRSpec.

\item
NIRSpec has detected emission in three (and perhaps four) Paschen transitions
and the He~I triplet at 1.083~\micron\ from TWA~27B, which are signatures 
of accretion when observed in young stars. 
These emission lines are the first evidence of
a circumstellar disk around TWA~27B. Using relations between Pa$\beta$ line 
luminosity and accretion luminosity from previous studies of stars and brown 
dwarfs \citep{nat04,alc17}, we have estimated an accretion rate of 
$\dot{M}\sim10^{-13}$--$10^{-12}$~$M_\odot$~yr$^{-1}$, which would be roughly 
consistent with the correlation between stellar mass and accretion rate for 
young stars and brown dwarfs \citep{muz05,her08,har16}.
TWA~27B does not exhibit excess emission in W2 (4.6~\micron) relative to other
young L-type objects. If it does have a disk, excess emission is likely 
present at longer IR wavelengths, which should be detectable with
5--28~\micron\ data from JWST/MIRI that were collected during
the same visit as the NIRSpec observations.

\item
We have calculated the bolometric luminosity of TWA~27B by integrating
its NIRSpec spectrum and estimating the missing flux at shorter and
longer wavelengths with model spectra, arriving at a value of 
log~$L/L_\odot=-4.466\pm0.014$.  Based on that luminosity and the age its 
association (10$\pm$2~Myr), TWA~27B should have a mass of 5--6~$M_{\rm Jup}$ 
according to evolutionary models \citep{bur97,cha23}.
These luminosity and mass estimates are similar to the values from
previous studies.

\end{enumerate}

\begin{acknowledgments}

P.T. acknowledges support from the European Research Council under grant
agreement ATMO 757858.  The JWST data were obtained from MAST at the Space 
Telescope Science Institute, which is operated by the Association of 
Universities for Research in Astronomy, Inc., under NASA contract 
NAS 5-03127. The JWST observations are associated with programs 
1270 and 1386. We acknowledge the team for program 1386 (PI: S. Hinkley) 
for developing their observing program with a zero-exclusive-access period.
This work used data from the ESA
mission Gaia (\url{https://www.cosmos.esa.int/gaia}), processed by
the Gaia Data Processing and Analysis Consortium (DPAC,
\url{https://www.cosmos.esa.int/web/gaia/dpac/consortium}). Funding
for the DPAC has been provided by national institutions, in particular
the institutions participating in the Gaia Multilateral Agreement. 
2MASS is a joint project of the University of
Massachusetts and IPAC at Caltech, funded by NASA and the NSF.
WISE is a joint project of the University of California, Los Angeles,
and the JPL/Caltech, funded by NASA. 
NEOWISE is a joint project of JPL/Caltech and the University of Arizona,
funded by NASA. This work used data from the
NASA/IPAC Infrared Science Archive, operated by JPL under contract
with NASA, and the VizieR catalog access tool and the SIMBAD database,
both operated at CDS, Strasbourg, France.
The Center for Exoplanets and Habitable Worlds is supported by the
Pennsylvania State University, the Eberly College of Science, and the
Pennsylvania Space Grant Consortium.

\end{acknowledgments}

\clearpage

\begin{deluxetable}{ll}
\tabletypesize{\scriptsize}
\tablewidth{0pt}
\tablecaption{Synthetic NIRSpec Photometry for TWA 27A/B, TWA 28, and VHS 1256b\label{tab:phot}}
\tablehead{
\colhead{Column Label} &
\colhead{Description}}
\startdata
Name & Source name \\
Jmag & $J$ 2MASS magnitude \\
Hmag & $H$ 2MASS magnitude \\
Ksmag & $K_s$ 2MASS magnitude \\
Jmkomag & $J$ MKO magnitude \\
Hmkomag & $H$ MKO magnitude \\
Kmkomag & $K$ MKO magnitude \\
W1mag & W1 WISE magnitude \\
W2mag & W2 WISE magnitude \\
m115mag & F115W NIRCam magnitude \\
m140mag & F140M NIRCam magnitude \\
m150mag & F150W NIRCam magnitude \\
m162mag & F162M NIRCam magnitude \\
m182mag & F182M NIRCam magnitude \\
m200mag & F200W NIRCam magnitude \\
m210mag & F210M NIRCam magnitude \\
m250mag & F250M NIRCam magnitude \\
m277mag & F277W NIRCam magnitude \\
m300mag & F300M NIRCam magnitude \\
m335mag & F335M NIRCam magnitude \\
m356mag & F356W NIRCam magnitude \\
m360mag & F360M NIRCam magnitude \\
m410mag & F410M NIRCam magnitude \\
m430mag & F430M NIRCam magnitude \\
m444mag & F444W NIRCam magnitude \\
m460mag & F460M NIRCam magnitude \\
m480mag & F480M NIRCam magnitude
\enddata
\tablecomments{
The table is available in its entirety in machine-readable form.}
\end{deluxetable}

\clearpage

\end{document}